\documentclass{elsart}
\usepackage{amsmath}
\usepackage{amsfonts}
\usepackage{amssymb}
\usepackage{bbm}
\usepackage{epsfig}

\begin{document}

\begin{frontmatter}
\title{Bivariate phase-rectified signal averaging}
   \author[ha]{Aicko Y. Schumann},
   %\thanksref{ca}},
   \author[ha]{Jan W. Kantelhardt},
   \author[mu]{Axel Bauer}
   and \author[mu]{Georg Schmidt}
   \address[ha]{Institut f\"ur Physik, Martin-Luther-Universit\"at, Halle, Germany }
   \address[mu]{Medizinische Klinik und Deutsches Herzzentrum der Technischen
   Universit\"at M\"unchen, Germany}
%   \thanks[ca]{Corresponding author: A. Y. Schumann, phone: +49 345 5525447;
%   fax: +49 345 5525446. {\it Email address:} aicko.schumann@physik.uni-halle.de}
\begin{abstract}

Phase-Rectified Signal Averaging (PRSA) was shown to be a powerful tool for the study
of quasi-periodic oscillations and nonlinear effects 
%destroying time-reversal invariance
in non-stationary signals.  Here we present a bivariate PRSA technique for the study
of the inter-relationship between two simultaneous data recordings.  Its performance
is compared with traditional cross-correlation analysis, which, however, does not
work well for non-stationary data and cannot distinguish the coupling directions in
complex nonlinear situations.  We show that bivariate PRSA allows the analysis of 
events in one signal at times where the other signal is in a certain phase or state; 
it is stable in the presence of noise and impassible to non-stationarities.

\end{abstract}
\end{frontmatter}

\section{Introduction}

Many natural systems generate periodicities on different time scales because some
of their components form closed regulation loops in addition to causal linear control
chains.  In biology and physiology, cardio-respiratory rhythms, rhythmic motions
of limbs in walking, rhythms underlying the release of hormones and gene expression,
membrane potential oscillations, oscillations in neuronal signals, and circadian
rhythms are just a few examples (see, e.g., \cite{Tyson2002,Glass2001}).  Oscillations
also occur in geophysical data, e.g., for the El-Ni\~no phenomenon, sunspot numbers,
and ice age periods \cite{Storch2001}. In many cases several signals from different
components of the complex system can be recorded simultaneously.  For understanding 
the control chains and loops in the system, we want to know how periodicities in the 
signals are generated by (possibly directed and/or nonlinear) interactions between 
its components.  Consequently, there is a need for identifying periodicities in one 
recorded signal together with the direction of causal relations to periodicities in 
other signals.  

Cross-correlation analysis and transfer function analysis are traditional tools for
this type of analysis.  However, there are three major drawbacks of these methods:
(i) only rather stationary data can be studied, (ii) a linear relationship between 
the signals is usually assumed, and (iii) the identification of causalities is 
hindered by the fact that the exchange of the two signals under study is identical 
with time inversion.  We thus propose a method which helps to overcome these problems.

Non-stationarities are a major problem in the analysis of signals recorded
from complex systems over a prolonged period of time
\cite{Priestly1988,Brockwell2003,Box1994,Kantz2004,Peng1994}.  Many internal
and external perturbations are continuously influencing the system and causing
interruptions of the periodic behavior.  The interruptions often 'reset' the
regulatory mechanisms resulting in phase de-synchronization of the
oscillations.  The signals thus become {\it quasi-periodic}, consisting of
many periodic patches as well as noise and trends.  Cross-correlation and
transfer function techniques are thus problematic.  In addition, there might
be causal inter-relations between two signals that cannot be revealed by these
methods.  For illustration, let us assume that a large increase {\it and} a
large decrease in signal $X$ (trigger signal) cause the same specific effect
in signal $Y$ (target signal), while there is no such effect in $Y$ if $X$
remains unchanged.  In this situation with an essentially nonlinear coupling
between the signals, both, cross-correlation analysis and spectral analysis
cannot reveal the effect.  They show the superposition of the two branches of
the interaction with opposite signs, i.e., no effect.  Even if the effects on
signal $Y$ were different for increases and decreases of signal $X$, one could
see some relation but could not distinguish the two effects.  Hence, one
needs a method that can separately study effects in signal $Y$ which might
occur in response to different causes in signal $X$, and vice versa.  A
separation of effects with different typical duration or frequency scale seems
also appropriate for distinguishing frequency-band selective inter-relations
between signals $X$ and $Y$.

Our approach for extracting inter-relations between two or more simultaneous
data recordings from a complex system is based on the phase-rectified signal
averaging technique (PRSA) \cite{PRSA_PhysicaA2006,PRSA_Lancet2006}, which was
shown to be a powerful tool for the study of quasi-periodic oscillations in
noisy, non-stationary signals.  The original method extracts the features in
one signal before and after increases in the same signal (or, alternatively,
decreases).  This way, information on characteristic quasi-periodicities,
short-term correlations, and time inversion asymmetry (causality) is
extracted, while non-stationarities and noise are eliminated.  The advanced
approach introduced here extracts the features in one signal before and after
increases in another signal.  Thus, the inter-relation between both signals
can be studied separately for both coupling directions, both time directions
and independent of non-stationarities and noise.

The paper is organized as follows. In Section 2 we describe both, the
univariate and the bivariate PRSA method.  Section 3 is dedicated to the
comparison of the bivariate PRSA method with the traditional cross-correlation
analysis.  We also address pitfalls and drawbacks of the cross-correlation
analysis that are often overlooked.  In Section 4 we discuss three model
examples and quantify the capacity of the bivariate PRSA method for the
detection of nonlinear interactions and quasi-periodicities. Finally, we
summarize and discuss possible applications in Section 5.

\section{PRSA methods}

\subsection{Univariate PRSA}

Let $X=(x_i)$, $i=1,\ldots, N$ be a long time series representing the signal
under investigation. In addition to periodicities and correlations of
interest, $X$ may contain non-stationarities, noise and recording artifacts.
One example for such a signal is the series of time intervals between
successive heartbeats determined from a long-term ECG (electrocardiogram) of a
patient in a hospital. Since the most pronounced peak in the ECG used for
heartbeat interval determination is called the R-peak, these time series are
often denoted as RR-interval (RRI) time series.  Phase Rectified Signal
Averaging was shown to reduce the signal to a much shorter sequence keeping
all relevant quasi-periodicities but eliminating non-stationarities,
artifacts, and noise \cite{PRSA_PhysicaA2006}. The PRSA algorithm consists of
three major steps as illustrated in Fig. 1.

\begin{figure}
   \begin{center}
      \epsfig{file=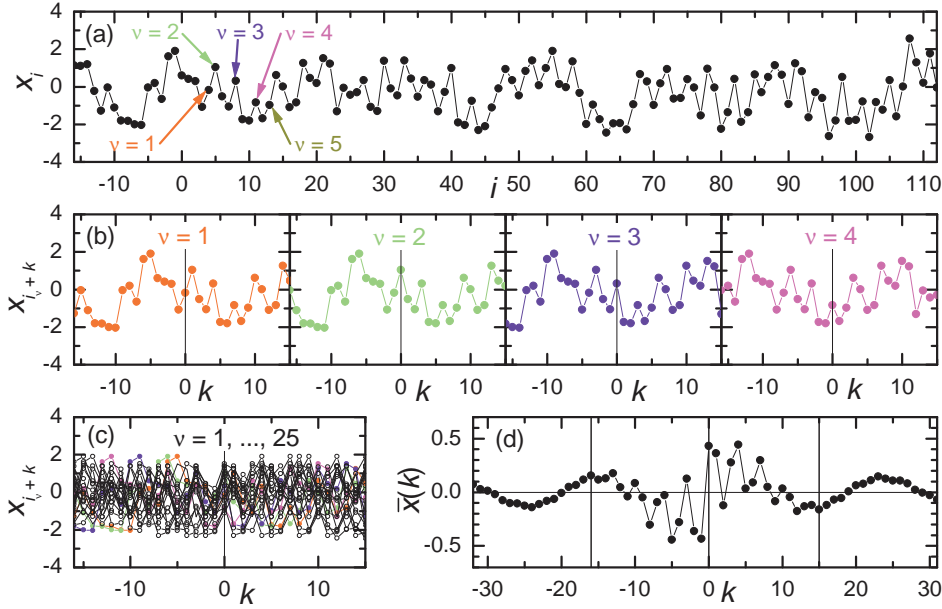,width=8cm,angle=-90}
   \end{center}
   \caption{Illustration of PRSA technique:  (a) Anchor points selected in the
   time series $(x_i)$, here: increases; only the first $5$ anchors are
   marked.  (b) Windows (surroundings, here: only first $4$ shown) of length
   $2L$ (here: $L=16$) defined around each anchor point.  (c) Surroundings of
   all anchor points on top of each other (here: only first $25$ shown).  (d)
   PRSA curve $\bar x(k)$ from averaging over all surroundings; the parameter
   $L$ is increased to $L=32$ in order to improve the visibility of the slow
   periodicity. The original signal is $1/f$ noise (generated with the Fourier
   filtering method) with two additional quasi-periodicities with
   characteristic frequencies $f = 0.05 /\Delta t$ and $f = 0.3 /\Delta t$;
   phase jumps are inserted after an average number of four periods (from
   \cite{PRSA_PhysicaA2006}).}
   \vspace{0.5cm}
   \label{fig1}
\end{figure}

{\it Step 1.} Anchor points in the time series are defined according to specific
features, e.g., increases (or, alternatively, decreases) in the time series 
(see Fig. 1(a)).  I.e., a point $x_i$ qualifies as an anchor point if
\begin{equation}
   x_i > x_{i-1}  \quad {\rm (or \; alternatively} \; x_i < x_{i-1}) \label{anchors}
\end{equation}
for triggering on increases or decreases, respectively.  In order to study
a lower frequency regime, averages of $T$ successive values of $x_i$ are compared
\cite{PRSA_PhysicaA2006}. Typically half of all points of the time series will
qualify as anchor points.  In general, quasi-periodic oscillations in a
noisy time series $X$ will result in anchor points predominantly found in
the phase of the steepest ascent (or decent for the second alternative in
Eq.~(\ref{anchors})), i.e., when the phase of the signal itself is close to $0$
(or close to $\pi$).  The phase information of the oscillations is thus obtained 
from the signal itself, and the signal can be {\it phase-rectified} using the 
anchor points. Note, that in principle any boolean valued function may be
used for the definition of anchor points, where true is associated with an
anchor while false is not. This allows studying more complex structures in signals.

{\it Step 2.} Windows, i.e., surroundings, of length $2L$ around each anchor point
$x_{i_\nu}$, $\nu=1,\ldots,M$, are identified (see Fig.~\ref{fig1}(b)); $M$ is 
the total number of regarded anchor points.  The surrounding of $x_{i_\nu}$ is
\begin{equation} x_{i_\nu-L}, x_{i_\nu-L+1}, \ldots, x_{i_\nu}, \ldots,
   x_{i_\nu+L-2}, x_{i_\nu+L-1}.
   \label{complete_windows}
\end{equation}
The parameter $L$ has to be chosen larger than the expected coherence time of the 
periodicities in the signal; it must definitely exceed the period of the slowest 
oscillation that one wants to detect.  All anchor points with indices $i_\nu$ 
smaller than $L+1$ and larger than $N-L+1$, i.e., at the very beginning and at 
the end of the time series, have incomplete surroundings.  The same holds for 
windows containing missing data points due to, e.g., measurement artifacts, 
instrument failure, or outliers.  

{\it Step 3.} All windows $\nu$, $\nu=1,\ldots,M$ are aligned at their anchor 
points $x_{i_\nu}$, and the phase-rectified signal average $\bar x(k)$ is obtained 
by averaging over all windows (see Figs.~1(c) and (d)),
\begin{equation}
   \operatorname{PRSA}_X(k) = \bar x(k) =\frac{1}{M} \sum_{\nu=1}^{M} x_{i_\nu+k} \text{,}
   \quad k=-L, \ldots, 0, \ldots, L-1.
   \label{PRSA_averaging}
\end{equation}
If $x_{i_\nu+k}$ is a missing data point, it is replaced by 0, and $M$ is
substituted by $M_k$ denoting the number of non-missing points at position $k$.
Including windows with missing data points yields better statistics and allows 
investigation of time series with a few artifacts. In general a well-behaved
average $\bar{x}(k)$ can be expected when there are at least $100$ to $1000$
anchor points, i.e., $N=200$ to $N=2000$ for the length of the record.

In the average \eqref{PRSA_averaging}, non-periodic components (not
phase-synchronized with the anchor points), i.e., non-stationarities,
non-identified artifacts, and noise, cancel out.  Only events that have a
fixed phase relationship with the anchor points, i.e., periodicities and
quasi-periodicities, 'survive' the procedure (see Fig.~\ref{fig1}).  The PRSA
signal $\bar x(k)$ represents the most important features of the original data
containing all quasi-periodicities aligned with phase zero in the center (at
$k=0$).  Applying the PRSA before traditional spectral analysis significantly
improves the quality of the spectra in the presence of noise and
non-stationarities \cite{PRSA_PhysicaA2006,PRSA_CHAOS2007}.  Differences
between PRSA curves obtained by applying either of the two criteria in
Eq.~(\ref{anchors}) will indicate missing time reversal symmetry of the
original signal.  Hence, nonlinear and non time-reversal invariant processes,
with different phenomena occurring during increasing and decreasing parts, can
be studied in detail.  Optionally, it might be meaningful to weight the
windows according to some criteria, e.g., according to the magnitude of
changes at anchor positions in the trigger signal. With anchors defined at
increases Eq.  \eqref{PRSA_averaging} becomes
\begin{equation}
   \operatorname{PRSA}_X(k) = \bar x(k) =\sum_{\nu=1}^{M} c_{i_\nu} x_{i_\nu+k} \text{,}
   \quad k=-L, \ldots, 0, \ldots, L-1\; .
   \label{PRSA_wheighted_averaging}
\end{equation}
with weights $c_{i_\nu}$, e.g.,
$c_{i_\nu}=(x_{i_\nu}-x_{i_\nu-1})/\sum_{\mu=1}^M(x_{i_\mu}-x_{i_\mu-1})$. Of
course, other weights could be defined as well.

\subsection{Bivariate PRSA (BPRSA)}

Now, we suggest a generalization of the univariate PRSA for studying the inter-relations
between two signals $X$ and $Y$.  If many signals are recorded simultaneously, representing 
the dynamics of the complex system, each pair can be characterized accordingly.  

\begin{figure}
   \begin{center}
      \includegraphics[width=\hsize]{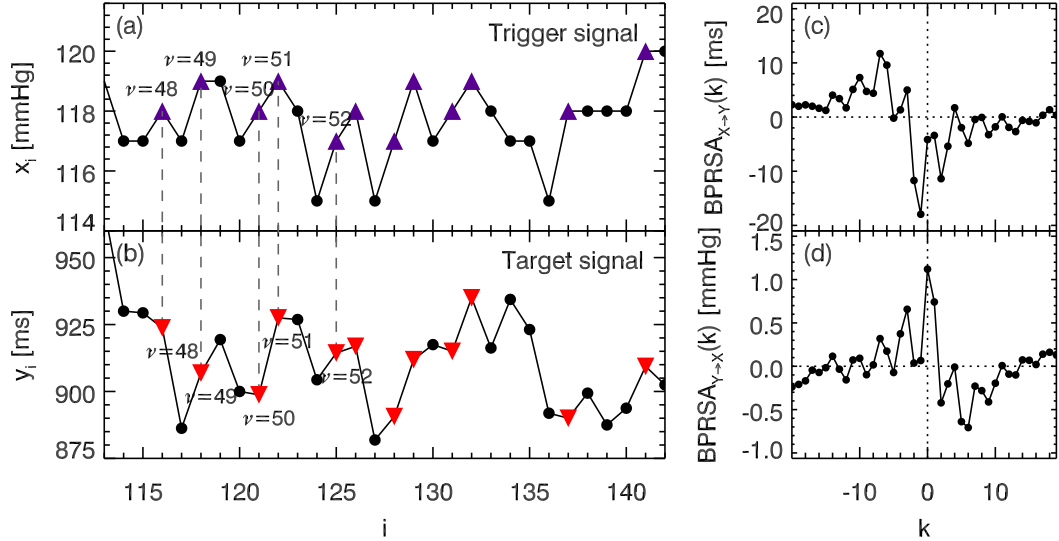}
   \end{center}
   \caption{Illustration of the BPRSA technique: According to
   Eq.~\eqref{anchors} anchor points are (a) selected in a trigger signal $X$
   and (b) transferred to the target signal $Y$. Here only parts of much
   longer blood pressure (a) and heartbeat interval (b) recordings are shown.
   After averaging the windows around each anchor point in $Y$ (red) according
   to Eq.~\eqref{BPRSA_averaging}, ${\rm BPRSA}_{X\rightarrow Y}$ is obtained
   (c). Likewise, changes in $X$ caused by increases in $Y$ can be studied by
   exchanging trigger and target signal, ${\rm BPRSA}_{Y\rightarrow X}$ in
   (d).} 
   \vspace{0.5cm}
   \label{BPRSA_figure}
\end{figure}

The method is nearly identical with the univariate approach described in the previous 
subsection, except for the usage of different signals in step one ($x_i=$ trigger signal 
$X$) and in steps two and three ($x_i=$ target signal $Y$).  Specifically, anchor points 
$i_\nu$, $\nu=1,\ldots,M$ are defined for increases (or alternatively decreases) in 
the trigger signal $X$, i.e. $(x_i)$ (step 1), while surroundings are defined (step 2) and 
averaged (step 3) for the target signal $Y$, i.e. $(y_i)$.  This yields the bivariate 
phase rectified signal average $\bar{y}(k)$:
\begin{equation} {\rm BPRSA}_{X \to Y}(k) = \bar y(k) = \frac{1}{M}
   \sum_{\nu=1}^{M} y_{i_\nu+k}, \quad k=-L, \ldots, 0, \ldots, L-1.
   \label{BPRSA_averaging}
\end{equation}
The transfer of the anchor points is illustrated in
Figs.~\ref{BPRSA_figure}(a),(b).

BPRSA is a non-symmetric algorithm, i.e., the exchange of trigger signal $X$
and target signal $Y$ will result in a different BPRSA curve, see
\ref{BPRSA_figure}(c),(d).  More complex
boolean or weighted anchor criteria, even ones based on more than one trigger
signal, are possible.  For example, the typical behavior of a target signal
$Y$ can be studied around points of time (anchors) with increases of signal
$X_1$ and positive values of signal $X_2$.  For a specific example of a
conditional anchor criterion consider the three signals heartbeat intervals,
respiratory phase, and blood pressure.  One can study characteristic heartbeat
intervals (target signal $Y$) around increasing systolic blood pressure (first
trigger signal $X_1$) and at a certain respiratory phase (second trigger
signal $X_2$).  First attempts revealed promising results for the
investigation of baroreflex properties and will be reported in a medical
journal.

\section{BPRSA and cross-correlation analysis}

\subsection{Cross-correlation analysis}

Although cross-correlation analysis is considered as a well established tool for
the study of inter-relations between two signals in many applications, only a few
authors have specifically studied its reliability
\cite{CCF_reliability_Peterson1998,CCF_reliability_Welsh1999,CCF_reliability_Vio2001}.
For two discretely measured signals $(x_{i})$ and $(y_{i})$, $i=1,\ldots,N$, the
{\it normalized cross-correlation function} is most commonly defined as
\begin{subequations}
\begin{eqnarray}
   \rho_{xy}(k) &=& \frac{1}{N \sigma_x \sigma_y} \sum_{i=1}^{N-k}
   (x_i-\mu_x) (y_{i+k}-\mu_y) \quad \text{for} \; k = 0, 1, \ldots \; \text{and} \\
   \rho_{xy}(k) &=& \frac{1}{N \sigma_x \sigma_y} \sum_{i=1-k}^{N}
   (x_i-\mu_x) (y_{i+k}-\mu_y) \quad  \text{for} \; k = -1, -2, \ldots.
\end{eqnarray}
    \label{discrete_CCF}
\end{subequations}

Here, $\mu_\alpha = \frac{1}{N} \sum_{i=1}^{N} \alpha_i$
and $\sigma_\alpha = \left[\frac{1}{N} \sum_{i=1}^{N} (\alpha_i - \mu_\alpha)^2
\right]^{1/2}$ are mean and standard deviation of both series $\alpha = x, y$,
respectively.  This definition assumes that both $\mu_\alpha$ and
$\sigma_\alpha$ do not vary in time, i.e., they do not depend on the segments
of the time series selected for the study.  This corresponds to the assumption
of weak stationarity. Strong stationarity additionally requires constancy of
all other moments.
For studies discussing the replacement of $\mu_x$ and $\mu_y$ by local estimates,
e.g. running averages, see \cite{Scargle1989,PressRybickiHewitt1992}.  Note,
however, that some cross-correlations might be reduced or eliminated by this
so-called pre-whitening procedure, which is therefore unsafe.

Another problem of cross-correlation functions is that the exchange of the two
signals $X$ and $Y$ corresponds to replacing $k$ by $-k$, i.e., time
inversion.  Hence, causality relations between the two series can hardly be
assessed. In general, the points of $\rho_{xy}(k)$ are highly auto-correlated,
e.g., $\rho_{xy}(k)$ is strongly correlated with  $\rho_{xy}(k+1)$. I.e.,
neighboring points in $\rho_{xy}(k)$ are stronger correlated with each other
than neighboring points in the original time series
\cite{CCF_reliability_Welsh1999,JenkinsWatts}.  This self-correlation causes
long living trends in $\rho_{xy}(k)$, e.g., a slow decay after a peak, which
is at risk of misinterpretation.

Furthermore, the sum in Eqs.~(\ref{discrete_CCF}) runs over $N-k$ terms, while
it is divided by $N$ instead of $N-k$.  This procedure corresponds to a
standard averaging procedure only in the limit of very long data ($N \to
\infty$).  Nevertheless, most statistical toolkits employ the definition
(\ref{discrete_CCF}), because the convolution theorem and fast Fourier
transform can be used to speed up the calculations significantly in this case
by application of the Wiener-Khinchin theorem.  Some authors even argue for an
increase in precision because the normalization $1/N$ reduces the mean-square
variance of $\rho_{xy}(k)$ (see, e.g., \cite{JenkinsWatts}).  However, this
non-matching prefactor results in a bias towards zero with increasing time lag
$k$ for small $N$, causing a triangular-shaped behavior of $\rho_{xy}(k)$.
Consequently, the value of $\vert k \vert >0$ for the center of a peak in
$\rho_{xy}(k)$ is systematically underestimated
\cite{CCF_reliability_Welsh1999}.  We are convinced that the correction factor
$N/(N-k)$ which transforms $\rho_{xy}(k)$ from Eqs.~(\ref{discrete_CCF}) into
the {\it correctly normalized cross-correlation function}
\begin{equation}
   \operatorname{CCF}_{X,Y}(k) = \frac{1}{(N-k) \sigma_x \sigma_y} \sum_{i=1}^{N-k}
   (x_i-\mu_x) (y_{i+k}-\mu_y),
   \label{discrete_CCF2}
\end{equation}
must be used to obtain reliable results except for very long data.

If the considered data is not fully stationary, one might want to use only the
values $x_i$ with $i=1,\ldots,N-k$ and $y_i$ with $i=k+1,\ldots,N$ for
calculating $\mu_x$, $\mu_y$, $\sigma_x$, and $\sigma_y$.  This approach is
known as {\it local cross-correlation} in literature; it is equivalent to the
Pearson $r_{xy}$ (product-moment) correlation coefficient for the two
overlapping pieces.  Since the partial means and standard deviations will
depend on $k$, the computational effort is significantly increased.  The bias
mentioned in the previous paragraph is not completely removed in this approach
\cite{CCF_reliability_Welsh1999} (although it is weaker than for the standard
definitions (\ref{discrete_CCF})).  In addition, problems with autoregressive
moving average processes (ARMA) were reported \cite{JenkinsWatts}.  Since the
cross-correlation approach does not work well for non-stationary data anyway,
we do not consider local cross-correlation here.

\subsection{Interpretation of BPRSA curves}

In BPRSA, anchor points usually occur in all parts of the trigger signal $X$.
The average of ${\rm BPRSA}_{X \to Y}(k) = \bar y(k)$ for all $k$ will thus be
approximately the global average of the whole signal, i.e., $\mu_y$.
Consequently, subtraction of this mean from $\bar y(k)$ yields positive and
negative values as in the cross-correlation function.  $\bar y(k)-\mu_y$ can
thus be interpreted in a similar way as an unnormalized cross-correlation
function.  If one divides by the global standard deviation, $\sigma_y$, the
resulting quantity
\begin{equation}
   {\rm BPRSA}_{X \to Y}^{\rm (norm)}(k) = \frac{\bar y(k)-\mu_y}{\sigma_y} 
   \label{norm-BPRSA}
\end{equation}
is also normalized. It can be compared with ${\rm CCF}_{X,Y}(k)$ in
Eq.~(\ref{discrete_CCF2}) and interpreted in a similar way.  Note that --
different from cross-correlation analysis -- this rescaling is just the last
step, and $\mu_y$ does not enter into the calculation of ${\rm BPRSA}_{X \to
Y}(k)$. Hence, the shape of the curve cannot be affected by
non-stationarities, i.e., inaccurate $\mu_y$. There is no practical advantage
of normalized BPRSA, unless the behavior of the curves for different signals,
e.g., triggering directions \fussy{$(X\to Y)$} and \fussy{$(Y \to X)$}, shall
be directly compared.  However, the global mean $\mu_y$ and global standard
deviation $\sigma_y$ might not exist due to non-stationarities and in this
case normalization is not recommended.

In some applications it is even preferred to study the unnormalized BPRSA
curves.  For example, in quantifying the action of blood pressure upon
heartbeat regulation via the baroreflex mechanism in the human cardiovascular
system, the variation of the time intervals between successive heartbeats in
reaction to increases in blood pressure needs to be measured.  In this case
the units of both signals have to be kept, and the measure characterizing the
baroreflex must have the unit ms/mmHg, i.e., time difference divided by
pressure difference.  In fact, cross-correlation studies can only yield either
quantities without units (if normalized) or quantities which are products of
both original units.  Quantities with the unit of only one original series or
their ratio (as needed for the baroreflex) cannot be obtained.  Hence, there
is no way to obtain a meaningful measure for the baroreflex from a
cross-correlation analysis, although the baroreflex is a typical example of a
meaningful inter-relation between two components of a complex system.

Effects occurring in ${\rm BPRSA}_{X \to Y}(k)$ for $k>0$ can be easily
recognized as consequences of the triggering events in the trigger signal $X$.
On the other hand, effects seen in ${\rm BPRSA}_{X \to Y}(k)$ for $k<0$ are
likely to be causes for the actual triggering events.  Note that a similar
conclusion is also valid for the cross-correlation function ${\rm CCF}_{X,Y}(k)$, since
effects observed for $k>0$ and $k<0$ are probably due to interactions from
signal $X$ onto signal $Y$ and vice versa.  However, BPRSA allows separating
these causality effects from nonlinear effects, as we will see in the
following.

%\sloppypar prevents formulas inside margin, more convenient than \linebreak
\sloppypar{Altogether, four BPRSA curves can be defined, compared with one
cross-correlation function:  ${\rm BPRSA}^{\nearrow}_{X \to Y}(k)$ (triggering
on increases in $X$), ${\rm BPRSA}^{\searrow}_{X \to Y}(k)$ (triggering on
decreases in $X$), ${\rm BPRSA}^{\nearrow}_{Y \to X}(k)$, and ${\rm
BPRSA}^{\searrow}_{Y \to X}(k)$ (triggering on $Y$). By comparing these
curves, additional information on the linearity of the interactions and time
reversal symmetry can be obtained. In the following we will use the symbols
$\nearrow$ and $\searrow$ for BPRSA curves obtained by triggering on increases
and decreases in the trigger signal only when necessary for distinction. In
all other cases ${\rm BPRSA}_{X\rightarrow Y}$ means ${\rm
BPRSA}^{\nearrow}_{X\rightarrow Y}$.}

If the interaction from signal $X$ to signal $Y$ is linear, we will find ${\rm
BPRSA}^{\nearrow}_{X \to Y}(k) = -{\rm BPRSA}^{\searrow}_{X \to Y}(k)$, since
increases and decreases in $X$ must cause opposite effects in $Y$.
Accordingly, ${\rm BPRSA}^{\nearrow}_{Y \to X}(k) = -{\rm BPRSA}^{
\searrow}_{Y \to X}(k)$ shows that the interaction from $Y$ to $X$ is linear.
If the interaction between both signals is fully symmetric, time inversion is
equivalent with exchanging the signals,  ${\rm BPRSA}^{\rm \nearrow(norm)}_{X
\to Y}(k) = {\rm BPRSA}^{\rm \nearrow(norm)}_{Y \to X}(-k)$ and ${\rm
BPRSA}^{\rm \searrow(norm)}_{X \to Y}(k) = {\rm BPRSA}^{\rm \searrow(norm)}_{Y
\to X}(-k)$.  Deviations from this behavior show non-symmetric coupling as do
deviations from ${\rm CCF}_{X,Y}(k)={\rm CCF}_{X,Y}(-k)$ in cross-correlation
analysis.  However, this can be checked independent of the linear or nonlinear
character of the interactions between the signals.  Note that normalized BPRSA
must be considered in this case, Eq. \eqref{norm-BPRSA}.  It is
straightforward to write down similar relations for testing further hypothesis
regarding the inter-relations between both signals.

\subsection{Comparison of cross-correlation analysis and BPRSA}

In this subsection we will see how the BPRSA overcomes the disadvantages of
cross-correlation analysis described before.

{\it 1. Causality and nonlinear interactions}.  As we have shown in the previous
subsection, more information on the linearity or nonlinearity of the interactions
and on time-reversal symmetry can be obtained from BPRSA curves than from the
cross-correlation function.

{\it 2. Time delays}.  The estimation of (positively or negatively) time-delayed
inter-relations between both signals is straightforward, just as in cross-correlation
analysis.

{\it 3. Missing data and outliers}.  BPRSA can easily cope with missing data 
(e.g., measurement artifacts, instrument failure, or outliers) in both series 
$X$ and $Y$.  Invalid values in $X$ just cannot become anchor points.  Invalid values 
in $Y$ will be disregarded (see text following Eq. \eqref{PRSA_averaging}). 

{\it 4. (Non-)stationarity of the data}.  In the definition of BPRSA (Section 2, in
particular Eq.~(\ref{BPRSA_averaging})) neither means nor standard deviations of
both signals $X$ and $Y$ are needed.  Hence, no direct problems arise for
non-stationary data.  In particular data with a piecewise constant trend, which is
often observed in medical data recordings, will cause no problems in BPRSA, because
Eq.~(\ref{BPRSA_averaging}) is a simple linear arithmetic averaging procedure.
The deviations from a small or large local average will have the same weight in
this averaging procedure.  Hence, BPRSA does not need pre-whitening of the data
before analysis.  Cross-correlation analysis, on the other hand, will be disturbed
severely by a piecewise constant trend, because the deviations $x_i - \mu_x$ from the
global average will be dominated by this trend (see Subsection 4.3 for an example).  
The same holds for an oscillating trend in the target signal $Y$ which is uncorrelated 
with the trigger signal $X$.  However, such a trend in $X$ will selectively cause anchor 
points and thus disturb also BPRSA; consequently more anchor points, i.e.,
longer data, will be needed!

A slowly varying, monotonous (e.g., polynomial) trend in the target signal will bend
the BPRSA curve, since the local means are different in the beginning and at the end
of the signal and in the beginning and at the end of each segment.  However, this bending
is definitely not stronger than a similar bending of the cross-correlation function.
Trends in the trigger signal $X$ will modify the fraction of anchor points for increases
and decreases, which has little effect on ${\rm BPRSA}_{X \to Y}(k)$ unless these trends 
are very strong.

{\it 5. Enhanced auto-correlations}.  Unlike the cross-correlation function
\cite{CCF_reliability_Welsh1999,JenkinsWatts}, which is often dominated by low
frequencies, BPRSA does not show artificially enhanced auto-correlations.  On
the contrary, low frequencies are reduced due to the filtering characteristics
(see next point).  This makes BPRSA particularly attractive for studying
signals with underlying $1/f$- rather than white noise. Note that $1/f$-noise
is prevalent, e.g., in medical and geophysical data.

\begin{figure}[t]
   \begin{center}
      \includegraphics[width=\hsize]{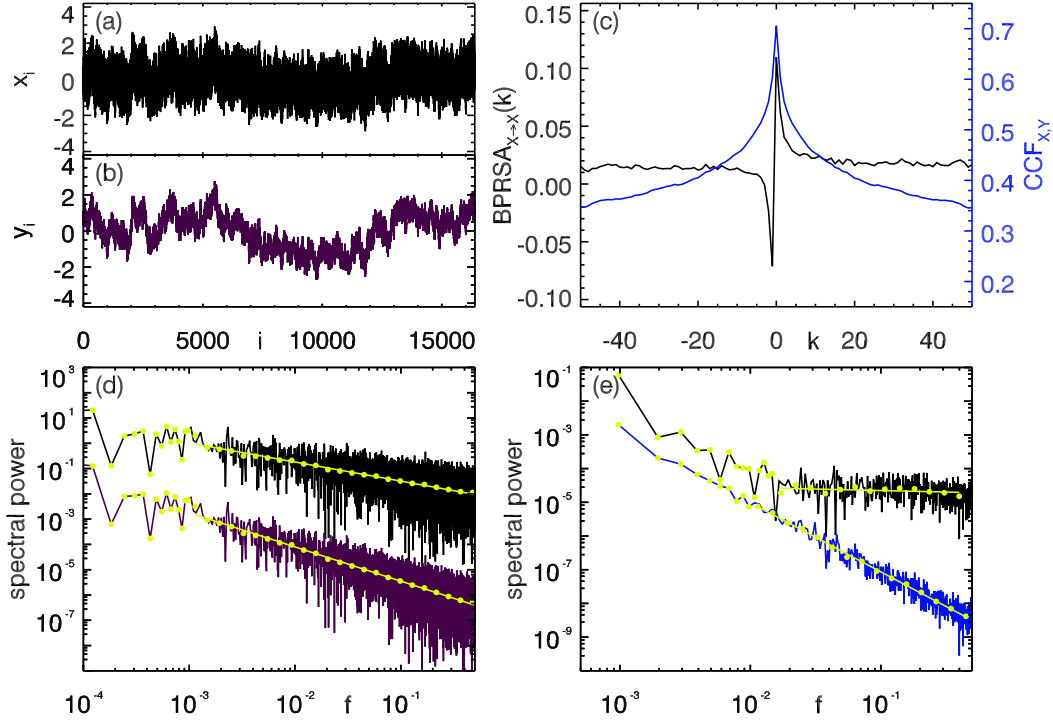}
   \end{center}
   \caption{Filter-properties of BPRSA and cross-correlation function for
   differently correlated noises with \sloppypar{$\mu_x=\mu_y=0$},
   \sloppypar{$\sigma_X=\sigma_Y=1$.} The time domain (a,b) and frequency
   domain (d) plots of $1/f^\beta$-noise with spectral exponents $\beta_X=0.7$
   (black curves, shifted) and $\beta_Y=1.3$ (magenta curves) are shown next
   to the $\operatorname{BPRSA}_{X\rightarrow Y}$ (black) and
   $\operatorname{CCF}_{X,Y}$ (blue) in (c) and their correctly normalized and
   correspondingly color-coded power spectra with
   $\beta_{\operatorname{BPRSA}_{X\rightarrow Y}}\approx 0$ (shifted) and
   $\beta_{\operatorname{CCF}_{X,Y}}\approx 2$ in (e). The two long-term
   correlated noises were generated by Fourier filtering with different
   $\beta$, starting both procedures with the same original white noise (not
   shown). For all spectra logarithmic binning and linear fitting (yellow dots
   and lines) were applied to estimate $\beta$.}
   \label{BPRSA_filter_properties}
\end{figure}

{\it 6. Filtering characteristics}.  Figure \ref{BPRSA_filter_properties}
compares the spectral properties of both, cross-correlation analysis and
BPRSA.  Since many interesting data contain long-term auto-correlations and
are characterized by $1/f$-noise in their power spectra, $P(f) \sim
f^{-\beta}$ with $\beta$ around 1, we start with two such noise series (see
Fig.~\ref{BPRSA_filter_properties}(a,b)) with $\beta_x \approx 0.7$ and
$\beta_y \approx 1.3$ (see Fig.~\ref{BPRSA_filter_properties}(d)).  The power
spectrum of the cross-correlation function decays as $f^{-2}$ (see
Fig.~\ref{BPRSA_filter_properties}(e)).  It is thus dominated by low-frequency
components.  The BPRSA curve, on the other hand, yields a nearly flat power
spectrum (see also Fig.~\ref{BPRSA_filter_properties}(e)). Therefore,
additional peaks and quasi-periodicities can be noticed and determined much
easier.

The filtering characteristics of BPRSA can be motivated as follows.  The
scaling behavior of the BPRSA spectrum is influenced by the anchoring
procedure in the trigger signal and by the averaging of the target signal.  We
want to estimate the probability $p(f)$ that an oscillating component with
frequency $f$, $y_f=A_y\sin(2\pi f t)$ in the target signal $Y$ affects ${\rm
BPRSA}_{X \to Y}(k)$ under the condition that an oscillation with the same
frequency $f$, $x_f=A_x\sin(2\pi f t)$ causes anchor points in the trigger
signal $X$.  Firstly, $x_f$ has to cause anchor points at positions $t_\nu$,
meaning $x_f(t_\nu)$ has to be larger than $x_f(t_\nu -\Delta t) \approx
x_f(t_\nu)-\Delta t x'_f = x_f(t_\nu)-\Delta t 2\pi f A_x \cos(2\pi f t_\nu)$
for anchor criterion Eq.~(\ref{anchors}a).  This is a valid approximation
except for very high frequencies $f$. The deviation $x_f(t_\nu)-x_f(t_\nu
-\Delta t)=\Delta t 2\pi f A_x\cos(2\pi f t_\nu)$ becomes maximal for
$t_\nu=n/f$ with any integer $n$.  Since anchor points $t_\nu$ are primarily
generated at or close to phase zero of the considered component $x_f$, the
later averaging is phase-rectifying in terms of the trigger signal. The value
of the maxima $x_f(t_\nu)-x_f(t_\nu -\Delta t)$ is $2\pi \Delta t f A_x$ and
thus the probability $p_x$ to anchor is proportional to $A_x f$.  On the other
hand, the component $y_f$ has an effect proportional to its amplitude $A_y$
due to the averaging procedure of Eq.~(\ref{PRSA_averaging}) and therefore
$p_y \sim A_y$.  The amplitude of the considered spectral components in ${\rm
BPRSA}_{X \to Y}(k)$ is thus determined by $A_x A_y f$.  If we consider two
signals $X$ and $Y$ consisting of correlated noise with power spectra
\begin{equation}
   P_x(f) \sim A_x^2 \sim f^{-\beta_X} \quad {\rm and} \quad
   P_y(f) \sim A_y^2 \sim f^{-\beta_Y}
\end{equation}
we obtain
\begin{equation}
   P_{\rm BPRSA}(f) \sim (p_x p_y)^2 \sim A_x^2 f^2 A_y^2 \sim
   f^{-\beta_X-\beta_Y+2} = f^{-\beta_{\rm BPRSA}}
\end{equation}

with $\beta_{\rm BPRSA} = \beta_X + \beta_Y - 2$, yielding $\beta_{\rm BPRSA}
\approx 0$ if both $\beta_X$ and $\beta_Y $ are close to one or their average
is close to one.
%The filter properties are best understood in terms of transfer function, or
%response function. The discrete Fourier transform of the cross-correlation is
%called the cross-spectral density or cross power-spectrum or just cross spectrum.

\section{Three illustrative examples}

Since BPRSA has significant advantages over cross-correlation analysis for
studying data with $1/f$ noise and/or nonlinear interaction as well as
non-stationary data, one can imagine several applications.  Here, we describe
three specific situations and illustrate the performance of BPRSA on model
data.

\subsection{White noises with linear relation}

\begin{figure}
   \begin{center}
      \includegraphics[width=\hsize]{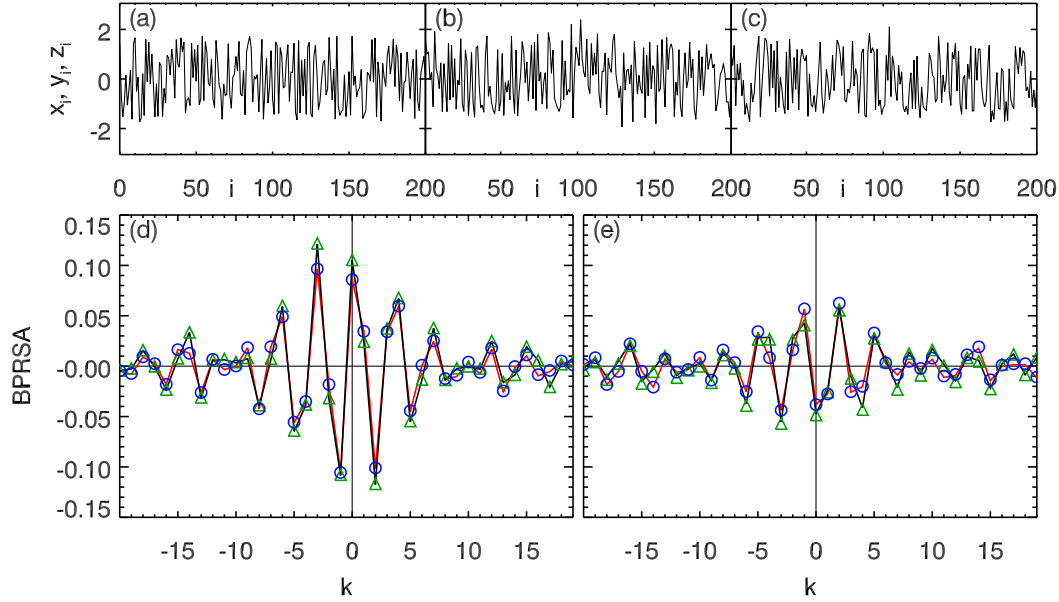}
   \end{center}
   \caption{Samples of the noise series $X$ (a, pure noise), $Y$ (b, generated
   from $X$ by Eq. \eqref{bandpass_filter} with $c_1=0.2$), and $Z$ (c,
   $c_1=-0.1$).  The HF band ($f\in [0.25,0.35]$ reciprocal sampling units) is
   used for the bandpass filtering, and the total length of the data is
   $N=16384$.  BPRSA results for $\alpha=Y$ (d) and $\alpha=Z$ (e):
   $\operatorname{BPRSA}^{\nearrow}_{X\rightarrow\alpha}$ (black solid lines), 
   $\operatorname{BPRSA}^{\nearrow}_{\alpha\rightarrow X}$ (red solid lines), 
   $-\operatorname{BPRSA}^{\searrow}_{X\rightarrow\alpha}$ (green triangles), 
   $-\operatorname{BPRSA}^{\searrow}_{\alpha\rightarrow X}$ (blue circles) 
   are shown. The points are connected for visual reasons only; all values are
   dimensionless.}
   \label{BPRSA_linear_relation}
\end{figure}

We consider two independent white noise signals $X=(x_i)$ and
$\tilde{Y}=(\tilde{y}_i)$ with zero mean and unit variance. Based on $\tilde{Y}$
we generate the signal $Y=(y_i)$ by introducing a linear unidirectional coupling with
$X$ in a certain frequency band. This is generated by calculating the linear
combination of $\tilde{Y}$ and one or more bandpass filtered components of
$X$,
\begin{equation}
   y_i = \tilde{y_i}+\sum_j c_j\operatorname{BP}^{(j)}_i(X).
   \label{bandpass_filter}
\end{equation}
The bandpass filtering is done in Fourier space, and $BP^{(j)}_i(X)$ denotes
the $i$-th element of the series obtained from the related $j$-th bandpass
filter operator acting on $X$. The prefactors $c_j$ include the coupling
strengths $\vert c_j \vert$ and directions ${\rm sgn}(c_j)$.  Finally, $Y$ is
normalized to obtain zero mean and unit variance.
Fig.~\ref{BPRSA_linear_relation}(a) illustrates the original noise $X$, while
Figs.~\ref{BPRSA_linear_relation}(b,c) show $Y$ and $Z=(z_i)$ for two
different values of $c_1$ and $c_j=0,\;\forall j>1$.

Different coupling strengths $\vert c_1 \vert$ are reflected by different
amplitudes of \linebreak $\operatorname{BPRSA}_{X\rightarrow\alpha}(k)$ and $\alpha
=Y, Z$, while a different coupling direction results in a different sign of
$\operatorname{BPRSA}_{X\rightarrow\alpha}(k)$ (compare
Figs.~\ref{BPRSA_linear_relation}(d,e)). Since we consider linear coupling,
$\operatorname{BPRSA}_{X\rightarrow\alpha}^{\nearrow}(k) =
-\operatorname{BPRSA}_{X\rightarrow\alpha}^{\searrow}(k)$ as discussed in
Section 3.2 and illustrated in Figs.~\ref{BPRSA_linear_relation}(d,e). There
is no advantage over ${\rm CCF}_{X,Y}(k)$ which looks very similar in this
example.

\subsection{Nonlinear relation}

\begin{figure}
   \begin{center}
      \includegraphics[width=\hsize]{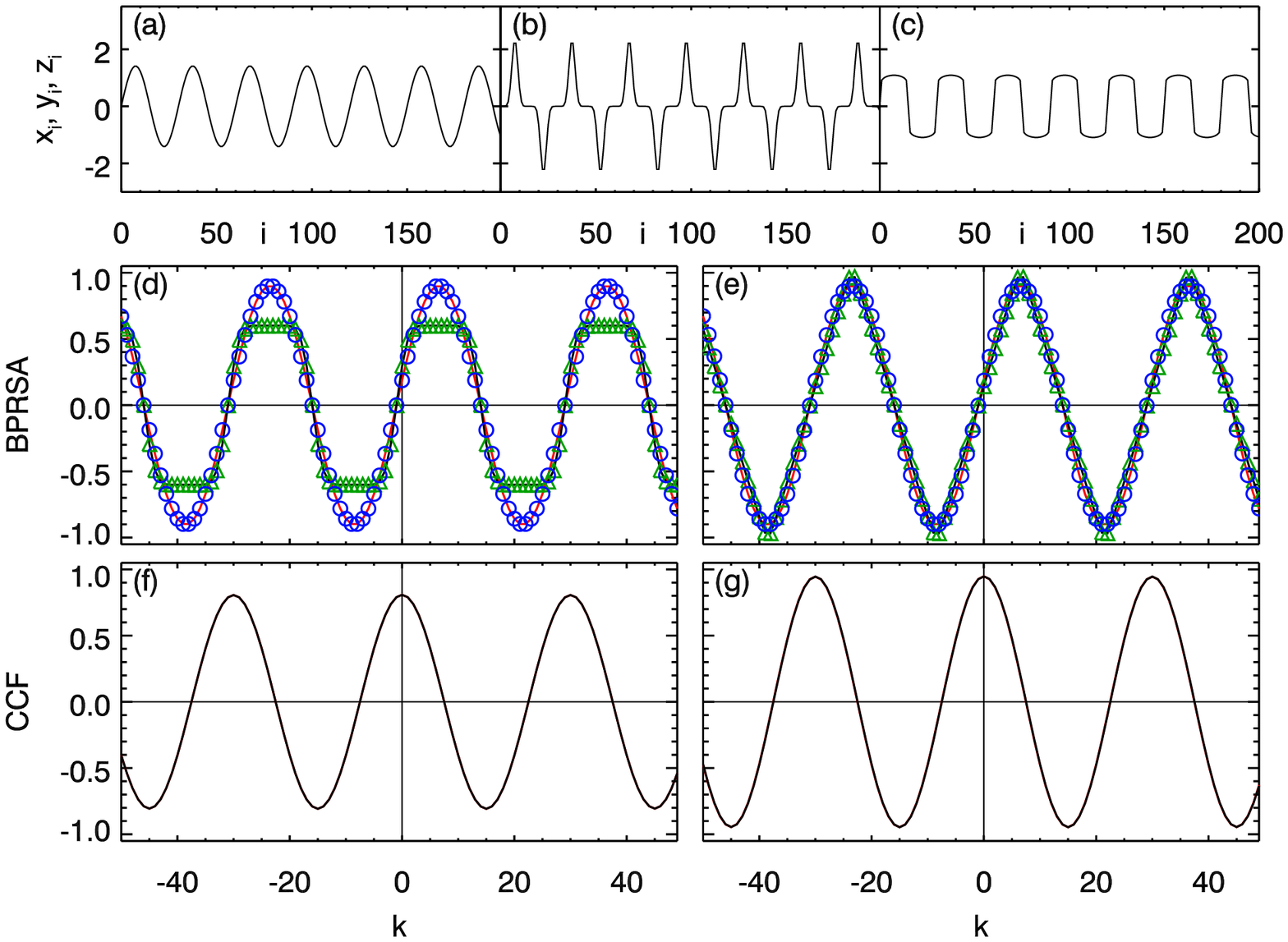}
   \end{center}
   \caption{(a) Sinusoidal signal $X$ and nonlinearly
   coupled signals (b) $Y$ and (c) $Z$ according
   to Eq. \eqref{nonlinear_signals}. (d)
   $\operatorname{BPRSA}^{\nearrow}_{X\rightarrow Y}$ (black solid lines),
   $\operatorname{BPRSA}^{\nearrow}_{Y\rightarrow X}$ (red solid lines), 
   $-\operatorname{BPRSA}^{\searrow}_{X\rightarrow Y}$ (green triangles),
   $-\operatorname{BPRSA}^{\searrow}_{Y\rightarrow X}$ (blue circles), (e)
   BPRSA for $Z$ instead of $Y$ accordingly; panel (f) shows
   $\operatorname{CCF}_{X,Y}=\operatorname{CCF}_{Y,X}$ and (g)
   $\operatorname{CCF}_{X,Z}=\operatorname{CCF}_{Z,X}$ (black on red solid
   lines)}
   \label{BPRSA_nonlinear}
\end{figure}

The response of the BPRSA to nonlinearly coupled trigger and target signals
strongly depends on the type of the coupling. The most simple nonlinear
coupling is the absolute value. Let us assume a sinusoidal trigger signal $X$
without noise and a target signal $Y$ that only contains the absolute value of
$X$, yielding a frequency doubling. When calculating the BPRSA all
oscillations cancel out and $\operatorname{BPRSA}_{X\longrightarrow
Y}(k)=\operatorname{BPRSA}_{Y\longrightarrow X}(k)=0$. In the presence of
additional $1/f$-noise the BPRSA will basically show features of the noise and
possibly finite size effects.  The same holds for similar nonlinear coupling,
e.g., raising to an even power. On the other hand, this elimination of higher
harmonics might be an advantage if one wants to clarify a complex relationship
between two unknown signals.

Now, we study nonlinear coupling without frequency doubling. Three simple
oscillating series are defined by
\begin{equation}
      x_i = \sin(2\pi f i),\quad
      y_i = (x_i)^9,\quad
      z_i = \operatorname{sgn}(x_i)\, \vert{x_i}\vert^{1/9}.
      \label{nonlinear_signals}
\end{equation}
and illustrated in Figs. \ref{BPRSA_nonlinear}(a-c).  The large power of $9$
has been chosen for visual reasons only; it enhances the differences as does
the absence of noise. The cross-correlation analysis (see Figs.
\ref{BPRSA_nonlinear}(f,g)) cannot distinguish (i) the cases $X\rightarrow Y$
and $X\rightarrow Z$ as well as (ii) both possible analysis directions.
Studying only the cross-correlation function could thus lead to the false
conclusion of equivalently related signals $Y$ and $Z$. BPRSA, on the other
hand, can clearly distinguish the four cases except for
$\operatorname{BPRSA}_{Y\rightarrow X}(k)=\operatorname{BPRSA}_{Z\rightarrow
X}(k)$. However, one has to keep in mind that the shape of the BPRSA curve
needs not be the same as the original target signal (compare Figs.
\ref{BPRSA_nonlinear}(b,d)).  A presence of noise might disturb the BPRSA
signal, making the identification of characteristics in trigger and target
signal more difficult, depending on the signal to noise ratio.

\subsection{Influence of Trends in the signal}

\begin{figure}
   \begin{center}
      \includegraphics[width=\hsize]{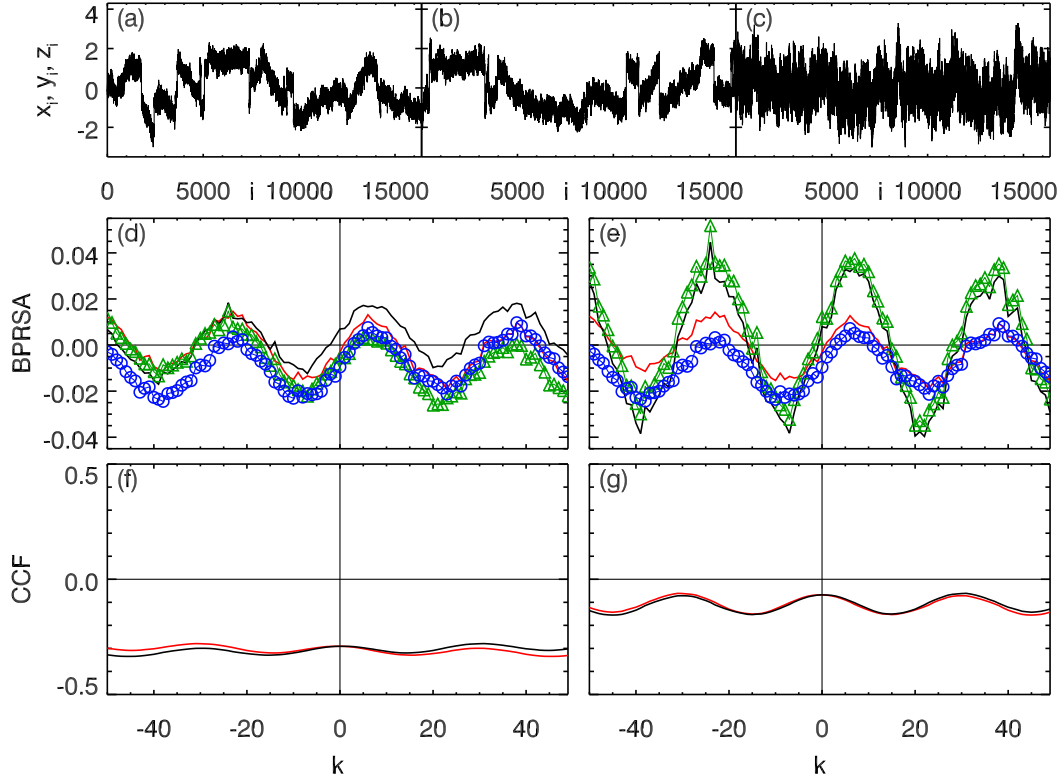}
   \end{center}
   \caption{Sinusoids with frequency $f=1/30$, amplitude $A=0.5$, and
   normalized additional $1/f$ noise with $\beta=1.0$; length $N=16384$. In
   (a,b), different partial trends of random offset, slope and duration were
   added; (c) is the same as (b) but without trends. (d) BPRSA results for
   $Y$: $\operatorname{BPRSA}^{\nearrow}_{X\rightarrow Y}$ (black solid
   lines), $\operatorname{BPRSA}^{\nearrow}_{Y\rightarrow X}$ (red solid
   lines), $-\operatorname{BPRSA}^{\searrow}_{X\rightarrow Y}$ (green
   triangles), $-\operatorname{BPRSA}^{\searrow}_{Y\rightarrow X}$ (blue
   circles).  (e) BPRSA for $Z$ replacing $Y$ accordingly. Panel (f) shows
   $\operatorname{CCF}_{X,Y}$ (black), $\operatorname{CCF}_{Y,X}$ (red) and
   (g) $\operatorname{CCF}_{X,Z}$ (black), $\operatorname{CCF}_{Z,X}$ (red)
   accordingly.  The points are connected for visual reasons only; all values
   are dimensionless.}
   \label{BPRSA_trends}
\end{figure}

Now let $X$ and $Y$ be two independent $1/f$-noise signals with zero mean and
unit variance generated by Fourier filtering. Furthermore, we add to both
signals a periodic component $A\sin(2\pi f i)$. Moreover, non-stationarities
are introduced by adding piecewise linear trends as follows. We start with
some initial value for the slope $a_1$ and the initial offset $a_0$. At random
positions, the offset and the slope are changed randomly within a previously
defined range; the trends added to $X$ and $Y$ are independent (see Figs.
\ref{BPRSA_trends}(a,b)). For comparison we define a third signal $Z$ that
equals $Y$ without trends (Fig.  \ref{BPRSA_trends}(c)).

Trends in the trigger signal will hardly affect the identification of the
anchor points, because the anchor criteria defined in Eq. \eqref{anchors} is
only based on local fluctuations. Note, that this might be different when
using a more sophisticated boolean anchor function as discussed earlier
(compare BPRSA directions $Y\rightarrow X$ and $Z\rightarrow X$ in Fig.
\ref{BPRSA_trends}(d,e)).

On the other hand, the influence of trends in the target signal cannot be
neglected (see Fig. \ref{BPRSA_trends}(e)).  In case of a significant global
trend in the target signal, e.g., more decreasing parts than increasing parts,
the global trend will be present in the BPRSA curve, although it is
diminished. Note, that due to trends which do not cancel out completely
$\operatorname{BPRSA}_{X\longrightarrow Y}^{\nearrow}(k)\ne
-\operatorname{BPRSA}_{X\longrightarrow Y}^{\searrow}(k)$ in
general (compare solid lines and triangles in Figs. \ref{BPRSA_trends}(d,e)).
When the BPRSA shows no trend at all, the target signal is either
characterized by no trends or the duration and slopes of increasing and
decreasing trends cancel out.

As an implication of the different influences of trends in the trigger and
target signal one can identify which signal is disturbed by trends by
comparing the BPRSA for opposite trigger-target directions ($X\rightarrow Y,\,
Y\rightarrow X$). This is inherently impossible with cross-correlation
analysis since the algorithm does not distinguish between both signals.
Besides, trends are harmful for the definition of a global mean and thus
disturb the standard cross-correlation analysis. Therefore, its results may
suggest a wrong correlation behavior. In Figs.~\ref{BPRSA_trends}(f),(g) one
finds, by chance, anti-correlated behavior although the signals themselves,
i.e., the sinusoids, are strongly positively correlated. For the same reason a
normalized BPRSA as defined in Eq.~\eqref{norm-BPRSA} cannot be applied here.
Of course, in this simple example the use of the local cross-correlation
function, which is based on local means rather than on a global mean, might
help to remove the influence of the trends.

\section{Summary and Outlook}

In summary we have shown that the BPRSA method has several advantages compared
with conventional cross-correlation analysis in the detection of
quasi-periodicities in noisy non-stationary data with oscillations of finite
coherence time. The method allows the analysis of the inter-relationship
between two signals, in particular effects in one signal triggered by events in
another signal.

This capability can be useful for the study of the inter-relation between
respiration, heart rate and blood pressure, i.e., the cardiovascular
regulation, which is an important topic in human physiology. Cardiovascular
functions are controlled by the tone of the sympathetic and parasympathetic
(autonomic) nervous system that is influenced by the baroreflex, a homeostatic
regulation that maintains a 'stable' blood pressure. An elevated blood
pressure reflexively causes the blood pressure to decrease and vice versa. It
is controlled through several stretch sensitive mechanoreceptors
(baroreceptors)\footnote{Activation of the baroreceptor results in an
inhibition of sympathetic components and activation of parasympathetic or
vagal components.  Due to an initially elevated blood pressure activated
baroreceptors tend to decrease cardiac output via a decrease in contractility
resulting in a lower heart rate and finally in a decrease in blood pressure. A
low blood pressure level relaxes the mechanoreceptor and stops the sympathetic
inhibition and results in an increased contractility, heart rate and blood
pressure.}. It is believed that cardiovascular illnesses disturb the
baroreflex. Related parameters might thus improve currently used predictors.
Hence, the detection of quasi-periodicities reflecting regulation processes of
the autonomic cardiac nervous system coinciding with increases or decreases
of blood pressure in long records of human heart rate is of high clinical
relevance.  Autonomic dysfunction is closely related to cardiac mortality and
susceptibility to life-threatening arrhythmic events \cite{Lown1976}.  The
assessment of heart rate variability by the PRSA based deceleration capacity
(DC) parameter \cite{PRSA_Lancet2006,PRSA_CHAOS2007} was shown to be superior
to spectral parameters proposed earlier for risk prediction \cite{Bigger1992}.
BPRSA seems to be promising for the definition of an advanced risk predictor
that respects the coupling of heart rate variability and blood pressure.

Further possible applications of BPRSA in biology and physiology include
rhythmic motions of limbs in walking, muscle contractions, rhythms underlying
the release of hormones that regulate growth and metabolism, periodicities in
gene expression, membrane potential oscillations, oscillations in neuronal
signals, and circadian rhythms \cite{Tyson2002,Glass2001}.  We believe that
the range of suitable applications for the BPRSA method also includes
quasi-periodic geophysical data, e.g., the El-Ni\~no phenomenon, sunspot
numbers, and ice age periods \cite{Storch2001}.  In addition, the analysis of
complex elastic wave patterns to study seismic events or to determine material
properties of granular matter might be improved by BPRSA.  The study of
non-stationary quasi-periodic complex waveforms is also a common task in the
analysis and recognition of speech or music.

{\it Acknowledgement:}
This study was supported by grants from the Deutsche Forschungsgemeinschaft
(grant KA 1676/3) and the European Union (project DAPHNet, grant 018474-2).

%\bibliographystyle{elsart-num}
%\bibliography{physjabb,references}

\begin{thebibliography}{10}
\expandafter\ifx\csname url\endcsname\relax
  \def\url#1{\texttt{#1}}\fi
\expandafter\ifx\csname urlprefix\endcsname\relax\def\urlprefix{URL }\fi

\bibitem{Tyson2002}
J.~Tyson, in: Computational Cell Biology: An Introductory Text on Computer
Modeling in Molecular and Cell Biology, J.~Tyson, J.~Wagner, E.~Marland,
C.~Fall (Eds.), Springer, New York, 2002.

\bibitem{Glass2001}
L.~Glass, Nature 410 (2001) 277.

\bibitem{Storch2001}
H.~von Storch, F.~Zwiers, Statistical Analysis in Climate Research, Cambridge
  University Press, 2001.

\bibitem{Priestly1988}
M.B.~Priestly, Nonlinear and Non-Stationary Time Series, Academic Press, New
York, 1988.

\bibitem{Brockwell2003}
P.J.~Brockwell, R.A.~Davis, Introduction to Time Series and Forecasting,
Springer, Berlin, 2003.

\bibitem{Box1994}
G.E.P.~Box, G.M.~Jenkins, G.C.~Reinsel,  
Time Series Analysis, Forecasting and Control, Prentice-Hall, Englewood Cliffs, 1994.

\bibitem{Kantz2004}
H.~Kantz, T.~Schreiber, Nonlinear Time Series Analysis, Cambridge University
Press, 2004.

\bibitem{Peng1994}
C.K.~Peng, S.V.~Buldyrev, S.~Havlin, M.~Simons, H.E.~Stanley, A.L.~Goldberger,
Phys. Rev. E 49 (1994) 1685.

\bibitem{PRSA_PhysicaA2006}
A.~Bauer, J.W. Kantelhardt, A.~Bunde, P.~Barthel, R.~Schneider, M.~Malik,
  G.~Schmidt, Physica A 364 (2006) 423.

\bibitem{PRSA_Lancet2006}
A.~Bauer, J.W.~Kantelhardt, P.~Barthel, R.~Schneider, T.M\"{a}kikallio,
  K.~Ulm, K.~Hnatkova, A.~Sch\"{o}mig, H.~Huikuri, A.~Bunde, M.~Malik,
  G.~Schmidt, Lancet 367 (2006) 1674.

\bibitem{PRSA_CHAOS2007}
J.W.~Kantelhardt, A.~Bauer, A.Y. Schumann, P.~Barthel, R.~Schneider,
  M.~Malik, G.~Schmidt, Chaos 17 (2007) 015112.

\bibitem{CCF_reliability_Peterson1998}
B.M.~Peterson, I.~Wanders, K.~Horne, S.~Collier, T.~Alexander, S.~Kaspi,
  D.~Maoz, PASP 110 (1998) 660.

\bibitem{CCF_reliability_Welsh1999}
W.~Welsh, PASP 111 (1999) 1347.

\bibitem{CCF_reliability_Vio2001}
R.~Vio, W.~Wamsteker, PASP 113 (2001) 86.

\bibitem{Scargle1989}
J.D.~Scargle, Astrophys. J. 343 (1989) 874.

\bibitem{PressRybickiHewitt1992}
W.H.~Press, G.B.~Rybicki, J.N.~Hewitt, Astrophys.
  J. 385 (1992) 404.

\bibitem{JenkinsWatts}
G.M.~Jenkins, D.G.~Watts, Spectral Analysis and Its Applications, 1st
  Edition, San Francisco: Holden-Day, 1969.

\bibitem{Lown1976}
B.~Lown, R.L.~Verrier, New Engl. J. Med. 294 (1976) 1165.

\bibitem{Bigger1992}
J.~Bigger, J.~Fleiss, R.~Steinman, L.~Rolnitzky, R.~Kleiger, J.~Rottman,
  Circulation 85 (1992) 164.

\end{thebibliography}

\end{document}